\documentclass[aps,prl,preprint,superscriptaddress]{revtex4-2}

\usepackage{graphicx}
\usepackage{siunitx}
\usepackage{xcolor}
\usepackage{bm}

\begin{document}


\title{Visualisation of quantised vortex reconnection as enabled by laser ablation}


\author{Yosuke Minowa}
\email[]{minowa@mp.es.osaka-u.ac.jp}
\affiliation{Graduate School of Engineering Science, Osaka University, 1-3, Machikane-yama, Toyonaka, Osaka, Japan}
\affiliation{JST, PRESTO, 4-1-8 Honcho, Kawaguchi, Saitama, Japan}

\author{Shota Aoyagi}
\affiliation{Graduate School of Engineering Science, Osaka University, 1-3, Machikane-yama, Toyonaka, Osaka, Japan}

\author{Sosuke Inui}
\affiliation{Department of Physics, Osaka City University, 3-3-138 Sugimoto, Osaka, Japan}

\author{Tomo Nakagawa}
\affiliation{Department of Physics, Osaka City University, 3-3-138 Sugimoto, Osaka, Japan}

\author{Gamu Asaka}
\affiliation{Department of Physics, Osaka City University, 3-3-138 Sugimoto, Osaka, Japan}

\author{Makoto Tsubota}
\affiliation{Department of Physics, Osaka City University, 3-3-138 Sugimoto, Osaka, Japan}
\affiliation{Nambu Yoichiro Institute of Theoretical and Experimental Physics (NITEP), Osaka City University, 3-3-138 Sugimoto, Osaka, Japan}
\affiliation{The Advanced Research Institute for Natural Science and Technology (OCARINA), Osaka City University, 3-3-138 Sugimoto, 558-8585 Osaka, Japan}

\author{Masaaki Ashida}
\affiliation{Graduate School of Engineering Science, Osaka University, 1-3, Machikane-yama, Toyonaka, Osaka, Japan}


\date{\today}

\begin{abstract}

Impurity injection into superfluid helium is a simple yet unique method with diverse applications, including high-precision spectroscopy\cite{fujisakiImplantationNeutralAtoms1993,takahashiSpectroscopyAlkaliAtoms1993,moroshkinLaserSpectroscopyPhonons2018}, quantum computing\cite{platzmanQuantumComputingElectrons1999}, nano/micro material synthesis\cite{moroshkinNanowireFormationGold2010,minowaInnerStructureZnO2017}, and flow visualisation\cite{guoVisualizationStudyCounterflow2010}. Quantised vortices are believed to play a major role in the interaction between superfluid helium and light impurities\cite{yarmchukObservationStationaryVortex1979,reifQuantizedVortexRings1964,takahashiSpectroscopyAlkaliAtoms1993,bewleySUPERFLUIDHELIUMVisualization2006,tangSuperdiffusionQuantizedVortices2021}. However, the basic principle governing the interaction is still controversial for dense materials such as semiconductor and metal impurities\cite{moroshkinMetallicNanowiresMesoscopic2016,moroshkinNanowireFormationGold2010}. Herein, we provide experimental evidence of the attraction of the dense silicon nanoparticles to the quantised vortex cores. We prepared the silicon nanoparticles via in-situ laser ablation. Following laser ablation, we observed that the silicon nanoparticles formed curved-filament-like structures, indicative of quantised vortex cores. We also observed that two accidentally intersecting quantised vortices exchanged their parts, a phenomenon called quantised vortex reconnection\cite{koplikVortexReconnectionSuperfluid1993}. This behaviour closely matches the dynamical scaling of reconnections. Our results provide a new method for visualising and studying impurity-quantised vortex interactions.
\end{abstract}


\maketitle

\section*{Introduction}
Bose–Einstein condensation is a remarkable manifestation of macroscopic quantum coherence, which does not entail any classical analogue, and has been the focus of intensive studies on fundamental quantum mechanics. Among the condensate types, superfluid $^4$He has a relatively higher transition temperature, allowing us to prepare a significantly larger number of superfluid atoms ($N \sim 10^{25}$). Thus, superfluid $^4$He serves as an eminent platform for studying the interaction between condensates and impurities including atoms\cite{takahashiSpectroscopyAlkaliAtoms1993}, ions\cite{reifQuantizedVortexRings1964}, electrons\cite{yarmchukObservationStationaryVortex1979}, and frozen hydrogen, deuterium and air particles\cite{bewleySUPERFLUIDHELIUMVisualization2006,tangSuperdiffusionQuantizedVortices2021,fondaReconnectionScalingQuantum2019}, because the introduced impurities immediately come into thermal equilibrium with the surrounding condensates, without destroying the superfluidity. A remarkable example is the interaction between the quantised vortex and light impurities such as electron bubbles\cite{yarmchukObservationStationaryVortex1979}. The quantised vortex is a stable topological defect that represents the macroscopic quantum nature of the superfluid $^4$He. The physics of the quantised vortex facilitates understanding of the fundamental properties of superfluid $^4$He. Owing to a Bernoulli pressure gradient, electron bubbles are attracted to the quantised vortex core, where they are stabilized. The trapped electron bubbles have been used to visualise the quantised vortex lattice\cite{yarmchukObservationStationaryVortex1979}. Similar interactions have been extensively studied for other light impurities (low mass density, low refractive index, $n\sim 1$). A stunning example is the visualization of quantised vortex dynamics using frozen hydrogen particles\cite{bewleySUPERFLUIDHELIUMVisualization2006, bewleyCharacterizationReconnectingVortices2008}. However, although some reports have indicated that dense materials such as metallic or semiconducting nanowires are formed along the quantised vortex core\cite{gordonRoleVorticesProcess2012,moroshkinNanowireFormationGold2010, moroshkinDynamicsVortexParticleComplexes2019,latimerPreparationUltrathinNanowires2014a}, the contribution of the quantised vortex remains controversial\cite{moroshkinMetallicNanowiresMesoscopic2016,moroshkinNanowireFormationGold2010}. Herein, we provide direct experimental evidence of dense silicon nanoparticle attraction to the quantised vortex, and the stabilisation along the vortex core. The density ($\rho_\mathrm{Si} \sim \SI{2400}{kg/m^3}$) and the refractive index ($n_\mathrm{Si}=4.14$ here) of silicon nanoparticles are distinctly larger than those of liquid helium ($\rho_\mathrm{He} \sim \SI{145}{kg/m^3}$, $n_\mathrm{He}=1.028$). Note that the optical refractive index roughly correlates with the mass density\cite{liuRelationshipRefractiveIndex2008}. The injection of the dense impurities into the superfluid $^4$He itself is not a straightforward task owing to the cryogenic environment. This is because the impurities are prone to stick to the substrate and aggregate, making it difficult to prepare the isolated impurities in the superfluid $^4$He. We overcame this problem by using a laser ablation technique, which allows the silicon nanoparticles to be prepared in the superfluid $^4$He. Laser ablation, or sputtering, is a unique in-situ preparation method to produce atoms, ions, and nano/microparticles in various environments; it is also applicable to superfluid $^4$He. Laser ablation in superfluid $^4$He has been used for precision spectroscopy\cite{fujisakiImplantationNeutralAtoms1993,takahashiSpectroscopyAlkaliAtoms1993,moroshkinLaserSpectroscopyPhonons2018}, tracer injection\cite{moroshkinImagingTimeresolvedStudy2020}, nano/micro structure formation\cite{moroshkinNanowireFormationGold2010,minowaInnerStructureZnO2017}, and optical manipulation/trapping\cite{inabaOpticalManipulationCuCl2006,minowaOpticalTrappingNanoparticles2021}. In this study, we observed the silicon nanoparticle-decorated, suspended quantised vortices, and identified the vortex reconnection events\cite{koplikVortexReconnectionSuperfluid1993}, the dynamics of which were found to be consistent with the results of dimensional analysis the simple vortex filament model (VFM) calculations.

\begin{figure}[h]
\includegraphics[width=0.7\columnwidth]{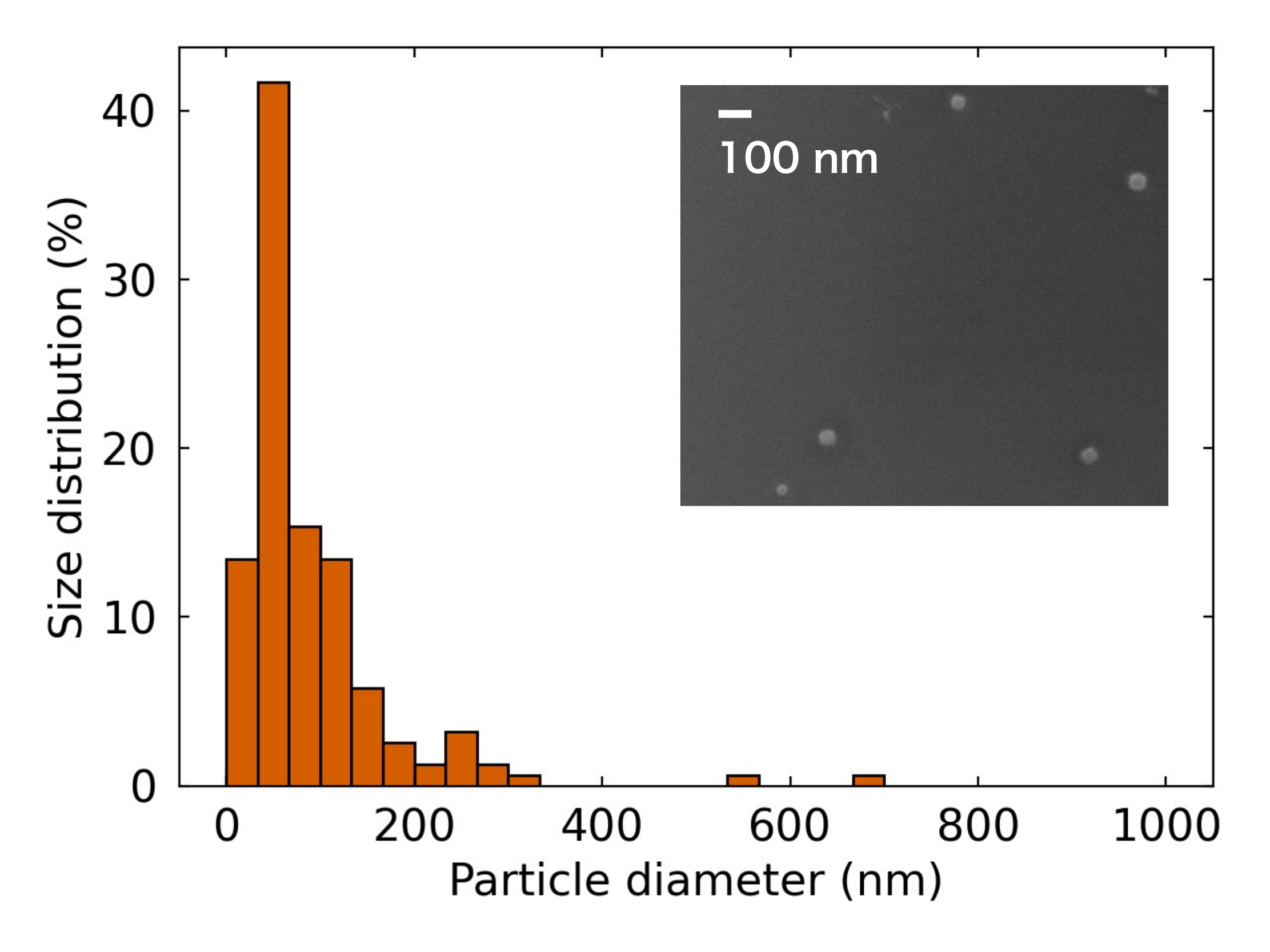}
\caption{\textbf{Silicon nanoparticles fabricated via laser ablation in superfluid $^4$He.} Silicon nanoparticle size distribution. An SEM image of typical silicon nanoparticles (inset).}
\label{fig:sem}
\end{figure}

\section*{laser ablation synthesis of nanoparticles and observation of suspended quantised vortex}
\begin{figure}[h]
\includegraphics[width=1.0\columnwidth]{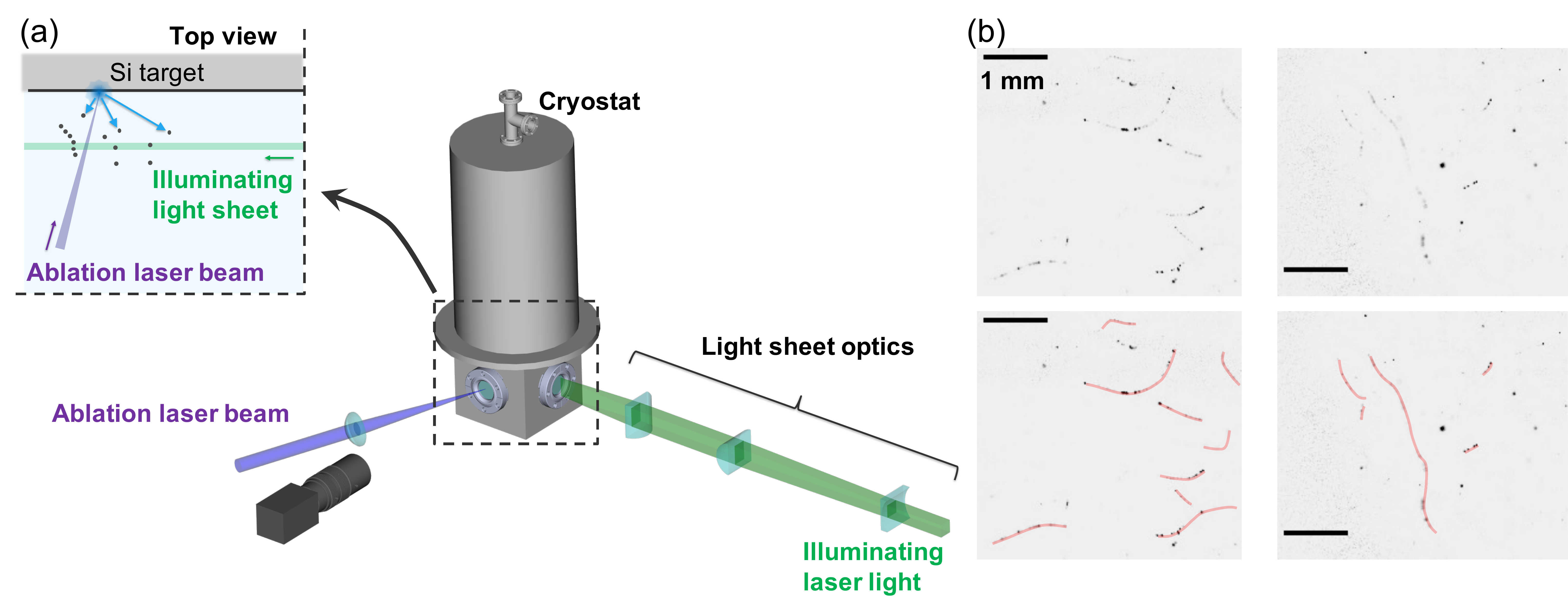}
\caption{\textbf{Observation of suspended quantised vortices decorated with silicon nanoparticles in superfluid $^4$He.} (a) Schematics of the experimental setup. Ablation laser pulses enter the cryostat from the front window and irradiate the silicon target. The nanoparticles stabilised along the quantised vortex core are illuminated from the side window with a light sheet. The scattered light is imaged through the front window onto a high-speed CMOS camera. The inset shows a schematic top view of the cryostat interior. Nanoparticles become dispersed after the laser ablation process. (b) Negative images of suspended quantised vortices visualised with silicon nanoparticles  (see also Supplementary Videos 1,2). Red lines indicate silicon nanoparticle structure (bottom). Scale bars, \SI{1}{mm}.}
\label{fig:setup}
\end{figure}

 We prepared semiconductor silicon nanoparticles via laser ablation technique at \SI{1.4}{K}. A single crystalline silicon target was placed in superfluid $^4$He and is irradiated with a laser light pulse. Subsequently, the melting and evaporation of the target and atom/ion/cluster ejection were initiated. Then, nanoparticles were formed following abrupt cooling of the ejected materials. The Fig. \ref{fig:sem} inset shows a scanning electron microscopy (SEM) image of the produced silicon nanoparticles observed at room temperature. The measured size distribution revealed that \SI{70}{\percent} of the particles are smaller than \SI{100}{nm} and \SI{50}{\percent} of the particles are smaller than $\SI{60}{nm}$. We observed the dynamics of the dispersed silicon nanoparticles under the illumination of a light sheet, as shown in Fig. \ref{fig:setup}(a) (see Methods for details). Figure \ref{fig:setup}(b) (top) shows typical negative (intensity inverted) images of the silicon nanoparticles suspended in superfluid $^4$He. Note that the small sizes of the nanoparticles relative to the spatial resolution of the applied optical system. Thus, the apparent size of the bright area should be interpreted as the scattering strength of the illuminating light, rather than the actual particle size. Many silicon nanoparticles were arranged in curved filaments that moved without disrupting the filament-like arrangement (see also Supplementary Videos 1, 2). The images in Fig. \ref{fig:setup}(b) (bottom) were modified to enable better visualisation of the filament-like structure locations. 

\section*{observation of quantised vortex reconnection}

\begin{figure}[h]
\includegraphics[width=1.0\columnwidth]{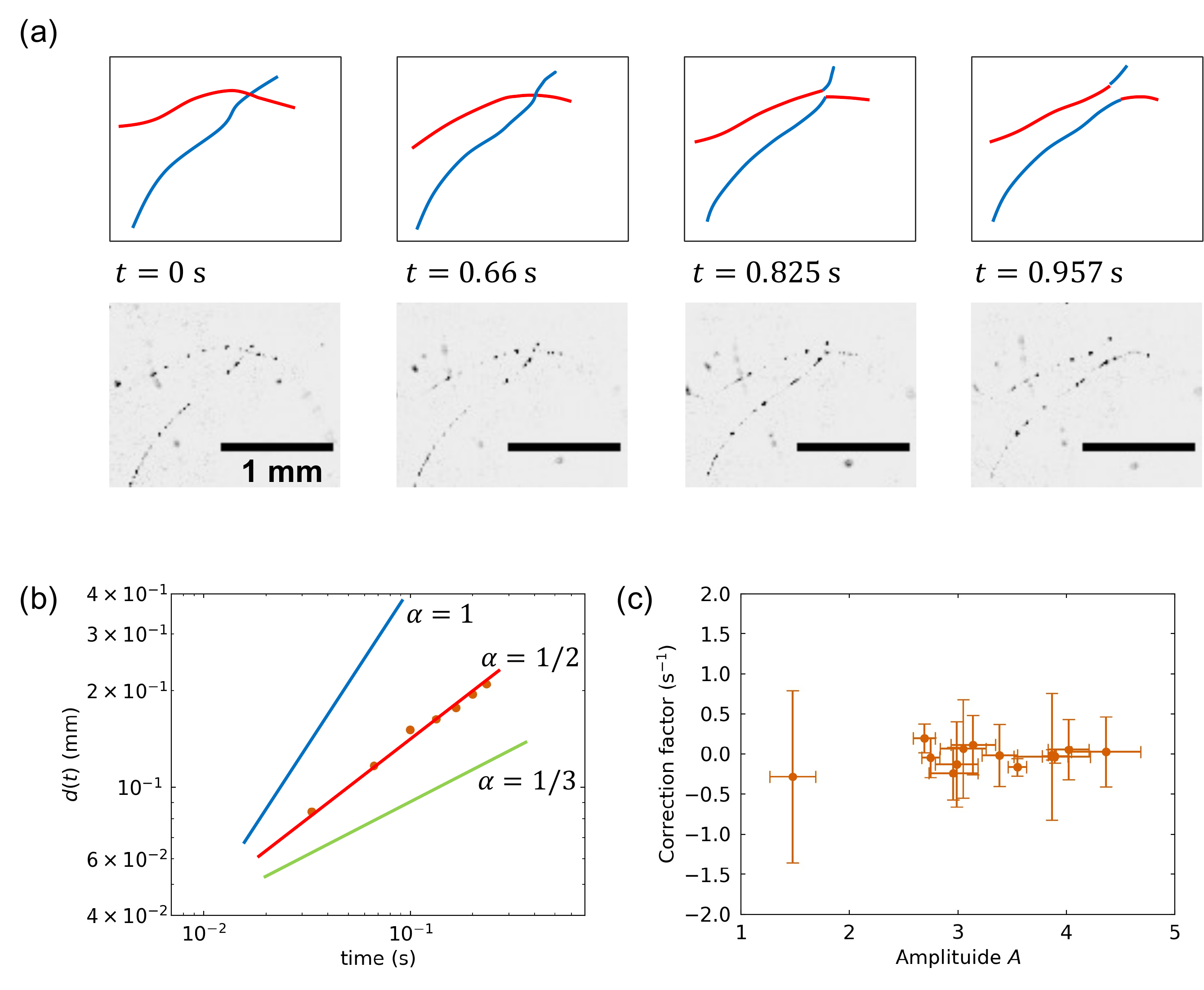}
\caption{\textbf{Reconnecting quantised vortices.} (a) Bottom row shows an image sequence of reconnecting quantised vortices, as captured with a high-speed CMOS camera. Top row shows the corresponding schematics of the reconnection. The moment of the reconnection was $t_0=\SI{0.79}{s}$. (b) Inter-vortex distance as a function of the time after the reconnection. Solid lines represent the power law, $d(t)\propto (t-t_0)^\alpha$ for different $\alpha$ values. (c) Distribution of fitted parameters: dimensionless amplitude $A$, and correction factor $C$. Each error bar shows the standard deviation resulting from the fitting.}
\label{fig:reconnect}
\end{figure}

The shape of the observed filament-like structures indicated that the loaded silicon nanoparticles were stabilized along the quantised vortex cores. Furthermore, we found filament reconnection events, which are one of the most characteristic phenomena of the quantised vortices\cite{bewleyCharacterizationReconnectingVortices2008,fondaDirectObservationKelvin2014}. When the two silicon-nanoparticle-decorated filaments intersected, there was an exchange, as shown in the bottom row of Fig. \ref{fig:reconnect}(a) and Supplementary Video 3. The corresponding schematic sequence is also shown in the top row of Fig. \ref{fig:reconnect}(a). Upon intersecting, the two filaments abruptly moved apart. The velocity and acceleration were clearly distinct from the slow background flow. These results demonstrate that this intersection and subsequent evolution are consistent with the theoretically predicted quantised vortex motion known as quantised vortex reconnection\cite{koplikVortexReconnectionSuperfluid1993}. If we assume that there was no characteristic length scale involved in the reconnection event, the inter-vortex distance $d$ (or any physical quantity having length dimension) after the reconnection can be expected to follow the dynamical scaling, given below:
\begin{equation}
    d(t) = A\sqrt{\kappa (t-t_0)}\label{eq:model}
\end{equation}
where $A$ is a dimensionless amplitude factor and $t_0$ is the moment of reconnection. This dynamical scaling has been previously observed\cite{fondaReconnectionScalingQuantum2019}. The equation was deduced from the dimensional analysis. The length dimension is only included in the circulation quanta $\kappa=h/m$, where $h$ is the Planck constant and $m$ is the mass of a helium atom. Thus, the temporal evolution of the inter-vortex distance should be written as equation \ref{eq:model}. Figure \ref{fig:reconnect}(b) depicts the inter-vortex distance measured for a pair of nearest nanoparticles in two reconnecting vortices after the moment of reconnection shown in Fig. \ref{fig:reconnect}(a). This figure also displays three different lines corresponding to the power-law $d(t)\propto (t-t_0)^\alpha$ for different $\alpha$ values. The close match between the experimental data and power law $d(t)\propto (t-t_0)^{0.5}$ indicates that the observed phenomenon is indeed the manifestation of the quantised vortex reconnection. 

To further elucidate the nature of the quantised vortex reconnection, we collected a number of similar events. Because of the wide depth-of-field of our applied optical system, the three-dimensional motion of the quantised vortex was projected onto the two-dimensional image. To prevent this from leading to the underestimation of the inter-vortex distance, we illuminated the suspending quantised vortices with a light sheet (see Methods for details). Then, the three-dimensional arrangement of the reconnecting vortices was only partly visualised. We focused on events wherein one of the reconnecting vortices was decorated and visualised with only one silicon nanoparticle. During the reconnection event, the nanoparticle collided with the visualised segment of the other quantised vortex; then, the magnitude and direction of the nanoparticle velocity suddenly changed, indicating the moment of the reconnection and following the repelling motion of the vortices. The measured inter-vortex distance can be fitted with the following equation, which includes the correction term:

\begin{equation}
    d (t) = A\sqrt{\kappa (t-t_0)}(1+c(t-t_0))\label{eq:model2}
\end{equation}
where $c$ is the correction factor. The correction considers localised environmental effect such as neighbouring vortices and boundary conditions\cite{paolettiReconnectionDynamicsQuantized2010}. The resulting fitted parameter distribution is shown in Fig. \ref{fig:reconnect}(c). As shown, the correction factors clustered around $c=0$, demonstrating consistency with the dimensional argument prediction. The dimensionless amplitude factor $A$ mainly spanned from $2.5 - 4.5$. Because the dimensional analysis did not provide any information about the value of $A$, we conducted a numerical simulation using a VFM. 

\section*{VFM calculations}
Several numerical studies have been conducted to reproduce the dynamics of quantised vortices in superfluid $^4$He.  One of the most successful approaches has been to treat a vortex line as a filament with an infinitesimal core size; this approach is known as the VFM (see Methods for more detail).
In the numerical simulations, we placed a pair of straight vortex lines in a cubic box of length  $1.0$ cm at $T = 1.4$ K.  To properly reproduce the actual experimental setup, we set the initial vortex configurations as follows: the vortices were initially separated by $d_0 = 0.1, 0.05 $ cm at their closest points and skewed to each other by some angle $\theta$ ($\theta$ = 0 for anti-parallel, and $\pi$ for parallel alignments, see Fig. 4 (a)--(d) for the numerical simulation with $d_0 = 0.1$ cm and $\theta = \frac{\pi}{3}$). Varying the initial angle $\theta$, we numerically estimated the closest distance $d$ as a function of time $t$ to verify the relation shown in equation (2).  Figure 4(e) clearly shows that the correction factor $c$ was relatively small for all amplitudes; this result is consistent with the experimental results shown in Fig. 3(c).  The numerically obtained range of the amplitude factor $A$ was 2.5 to 4.5, which is also in good agreement with the experimental results. Thus, we can conclude that the characteristic motions of the aggregated silicon particles indeed stems from the reconnection events of a pair of quantised vortices.

\begin{figure}[h]
\includegraphics[width=1.0\columnwidth]{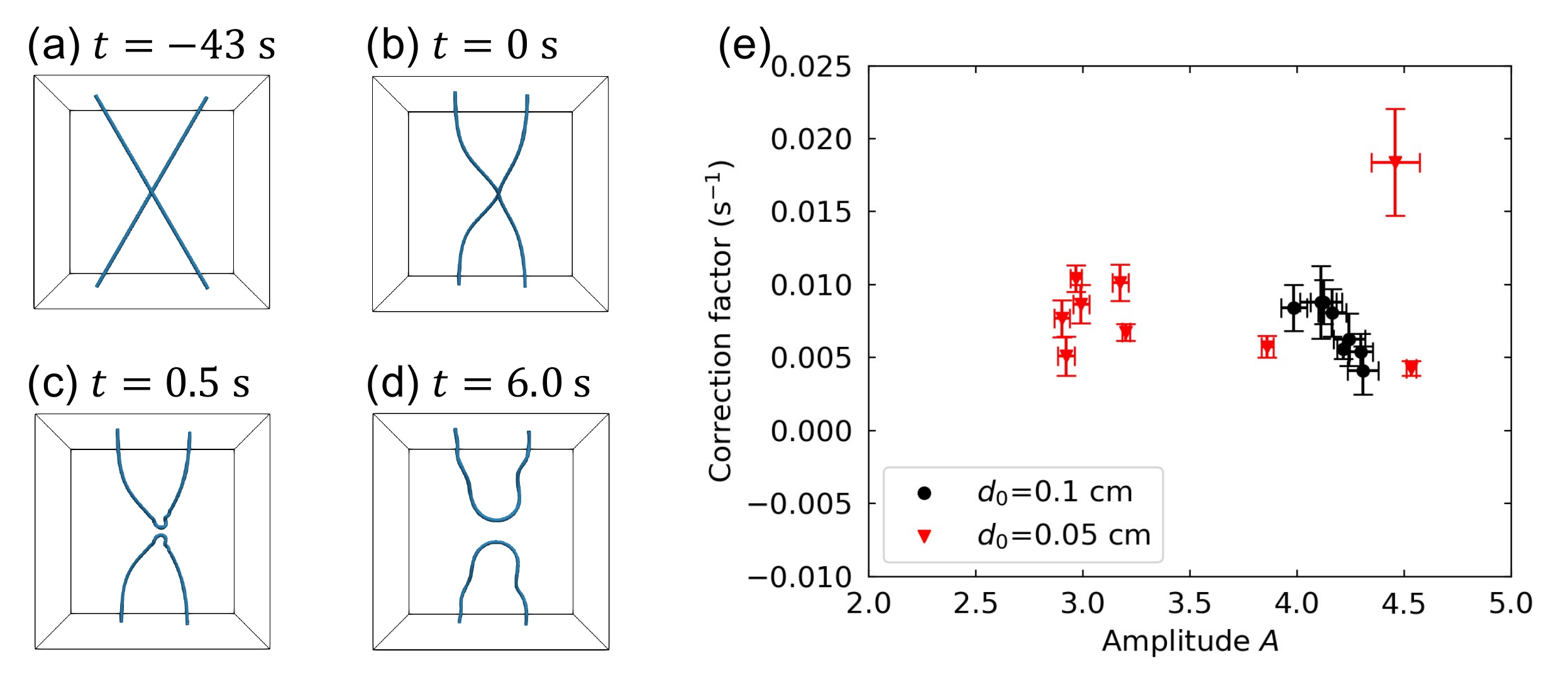}
\caption{\textbf{Calculation of reconnecting quantised vortex filaments.} (a)-(d) Snapshots of simulated reconnection of quantised vortex filaments. Each cube length was $1.0$ cm, and the top and bottom surfaces were subject to the solid boundary condition. (e) Distribution of fitted parameters: dimensionless amplitude $A$ and correction factor $c$ for the simulated results. Each scattered data point corresponds to a single set of simulation with some initial angle $\theta \in [0,\pi/2)$, and the each error bar shows the standard deviation resulting from applying least-squares fitting to the numerical results of equation. (2).}
\label{fig:simulation}
\end{figure}

\section*{Discussion and outlook}
In this study, we demonstrated that the dense silicon nanoparticles are stabilised along the quantised vortex cores and move collectively with the quantised vortex. This finding clearly supports the hypothesis that the quantised vortex plays a major role in the formation of the semiconducting or metallic nanowires in superfluid $^4$He\cite{moroshkinNanowireFormationGold2010,gordonRoleVorticesProcess2012,latimerPreparationUltrathinNanowires2014a,spenceVortexinducedAggregationSuperfluid2014,thalerFormationBimetallicCoreshell2014,volkThermalInstabilitiesRayleigh2015,moroshkinMetallicNanowiresMesoscopic2016}, although the intermediate process needs to be further studied. The laser ablation method can inject various nanoparticles, ranging from semiconductors\cite{inabaOpticalManipulationCuCl2006} to metals\cite{moroshkinNanowireFormationGold2010, minowaOpticalTrappingNanoparticles2021}, into superfluid $^4$He. This feature of material diversity is important for the integration of quantised vortex research and optical probing and optical manipulation, because it enables selection of an appropriate material based on favourable optical properties. The refractive index of silicon ($n_\mathrm{Si}=4.14$) is indeed much larger than that of frozen hydrogen ($n_\mathrm{H2}=1.14$)\cite{fondaSubmicronSolidAir2016}, a tracer typically used in superfluid $^4$He. This high refractive index allows the light to be scattered with high efficiency, leading to a larger signal-to-noise ratio and higher-speed observation. For the same size of the nanoparticles, the scattering cross section of silicon nanoparticles was larger than that of solid hydrogen by two orders of magnitude (see Supplementary Information). The higher scattering efficiency also allowed us to use smaller particles as the quantised vortex tracer, as it ensured a more passive role for the tracer. In fact, the size of the silicon nanoparticles used in this study was smaller than that of previously reported solid hydrogen (a few micrometre scales\cite{bewleyCharacterizationReconnectingVortices2008}) and solid atmospheric air (\SI{400}{nm}\cite{fondaSubmicronSolidAir2016}). The flexibility of the laser ablation technique allowed us to control the size of the nanoparticles by changing the laser ablation parameters such as pulse duration, wavelength, and pulse energy\cite{semaltianosNanoparticlesLaserAblation2010}. Various material options and size controllability are significant advantages that provide an alternative method not only for the quantised vortex visualization, but also for liquid helium flow visualization. Moreover, these features are important for the incorporation of optical manipulation techniques\cite{inabaOpticalManipulationCuCl2006,minowaOpticalTrappingNanoparticles2021} into quantum fluid research.

\section*{methods}

\subsection*{Silicon nanoparticle injection and quantised vortex visualization in superfluid $^4$He}
We placed a \SI{3}{cm} x \SI{3}{cm} x \SI{3}{cm} cuvette in superfluid $^4$He. The entire experimental process was performed in this cuvette filled with the superfluid $^4$He. We mounted a single crystalline silicon target in the cuvette. The liquid helium temperature was maintained at approximately 1.4 K throughout the experiment. Nanosecond light pulses from a frequency-tripled Q-switched Nd:YAG laser (wavelength: 355 nm, pulse duration: 10 ns, repetition rate: 10 Hz, and pulse energy: 1 mJ) were focused onto the target surface with a spot size of $\sim$\SI{40}{\micro \meter} using a lens with 200-mm focal length. The produced silicon nanoparticles were illuminated using a continuous-wave laser (wavelength: 532 nm and power: 1 W). The illuminating light was in the form of a laser sheet (full width: \SI{300}{\micro \meter}), prepared by using a set of cylindrical lenses. The spatial resolution of our observation system ($\SI{19}{\micro \meter}$) was primarily limited by the CMOS camera pixel size. All data were collected at 30 frames per second. The images for the quantised vortex observation were intensity-inverted and background-subtracted. Then, the contrast was enhanced to clarify the observed structures. 

\subsection*{Vortex filament calculations}
Superfluid $^4$He can be described by a two-fluid model, $\it{i.e.}$ it comprises an inviscid superfluid component and a viscous normal fluid component.  Any rotational motion in superfluid flow is sustained by quantised vortex filaments \cite{donnellyQuantizedVorticesHelium1991}.
According to Helmholtz’s theorems, a vortex filament $\bm{s}(\xi,t)$ travels with a superfluid velocity $\bm{v}_{s0}(\xi,t)$ induced at its location, where $\xi$ is its arc length. The velocity can be found by calculating a Biot-Savart integral of the form given below\cite{schwarzThreedimensionalVortexDynamics1985}:
\begin{equation}
		\bm{v}_{\mathrm{s}0}(\bm{r}) = \frac{\kappa}{4\pi} \int_\mathcal{L} \frac{  \bm{s} ^{\prime}(\xi) \times  (\bm{s}(\xi)-\bm{r})   }{| \bm{s}(\xi)-\bm{r} |^{3}} d\xi,
\end{equation}	where the prime symbol represents a derivative with respect to $\xi$ and $\mathcal{L}$ is the path that coincides with the vortex filament.  At finite temperature, however, the motion of a vortex line can be modified by the so-called mutual friction that is ascribed to its scattering of normal fluid component, as follows:
\begin{equation} \label{eq: Schwartz}
\frac{d	\bm{s}(\xi,t)}{dt}= \bm{v}_{s0}(\xi,t)  + \alpha   \bm{s}'\times(\bm{v}_\mathrm{n}-\bm{v}_{\mathrm{s}0}) 
		 +\alpha' \bm{s}' \times [ \bm{s}' \times (\bm{v}_\mathrm{n} - \bm{v}_{\mathrm{s}0}) ],
\end{equation}
where $\alpha$ and $\alpha'$ are temperature-dependent mutual friction coefficients, and $\bm{v}_\textrm{n}$ is some background normal fluid flow.  The position of a filament $\bm{s}(\xi)$ was discretised and stored as a set of points separated by some resolution $\Delta \xi$. 
Computationally, the spatial resolution $\Delta \xi$ was set to be within a range of $\Delta \xi_\textrm{min} = 0.05$ mm to $\Delta \xi_\textrm{max} = 0.1$ mm, and the temporal resolution $\Delta t$ is $0.01$ s. 
We, then, solved the integro-differential equation (equation (\ref{eq: Schwartz})) adopting a fourth-order Runge--Kutta integration method for the temporal evolution. 

Vortex reconnection events can be handled algorithmically in VFM. When two vortices were within $\Delta \xi_\textrm{min}$ of each other, we exchanged the legs of vortices.  This method may seem arbitrary; however, we were primarily interested in the post-reconnection dynamical scaling behaviour in vortices, and the detail of algorithm does not significantly affect the scaling law.

\section*{Acknowledgements}This work was supported by the MEXT/JSPS KAKENHI grant numbers JP20H01855, JP20J23131, JP16H06505, and JP18KK0387, and by JST, PRESTO grant number JPMJPR1909, Japan.

\section*{Author contributions}
\subsection*{Contributions}Y.M. conceived and designed the experiment. Y.M. and S.A. performed the experiments and analysed the data. S.I., T.N., G.A., and M.T. conducted the vortex filament simulation and the analysis. Y.M. and M.T. wrote the paper. M.A. provided technical support, and helped write and edit the manuscript.
\subsection*{Corresponding author}
Correspondence and requests for materials should be addressed to Yosuke Minowa.

\section*{Ethics declarations}
\subsection*{Competing interests}
The authors declare no competing interests.

\section*{Additional Information}
Supplementary Information is available for this paper. 

\bibliography{00ablationvortex}

\end{document}